\input{aipcheck}

\documentclass[
    ,final            
  ]
  {aipproc}

\layoutstyle{6x9}

\begin{document}

\title{New insights on the accretion disk-winds connection in radio-loud AGNs from Suzaku}


\author{F. Tombesi}{
  address={CRESST, NASA/GSFC/CRESST, Greenbelt, MD 20771, USA}
, altaddress={Department of Astronomy, University of Maryland, College Park, MD 20742, USA}
}

\author{R. M. Sambruna}{
  address={George Mason University, MS 3F3, 4400 University Drive, Fairfax, VA 22030}
}

\author{J. N. Reeves}{
  address={Keele University, Keele, Staffordshire ST5 5BG, UK}
}

\author{V. Braito}{
  address={University of Leicester, University Road, Leicester LE1 7RH, UK}
}

\author{M. Cappi}{
  address={INAF-IASF Bologna, Via Gobetti 101, I-40129 Bologna, Italy}
}

\author{C. S. Reynolds}{
  address={Department of Astronomy, University of Maryland, College Park, MD 20742, USA}
}

\author{R. F. Mushotzky}{
  address={Department of Astronomy, University of Maryland, College Park, MD 20742, USA}
}


\begin{abstract}
From the spectral analysis of long \emph{Suzaku} observations of five radio-loud AGNs we have been able to discover the presence of ultra-fast outflows with velocities $\sim$0.1c in three of them, namely 3C~111, 3C~120 and 3C~390.3. They are consistent with being accretion disk winds/outflows. 
We also performed a follow-up on 3C~111 to monitor its outflow on $\sim$7~days time-scales and detected an anti-correlated variability of a possible relativistic emission line with respect to blue-shifted Fe K features, following a flux increase. This provides the first direct evidence for an accretion disc-wind connection in an AGN.
The mass outflow rate of these outflows can be comparable to the accretion rate and their mechanical power can correspond to a significant fraction of the bolometric luminosity and is comparable to their typical jet power. Therefore, they can possibly play a significant role in the expected feedback from AGNs and can give us further clues on the relation 
between the accretion disk and the formation of winds/jets.

\end{abstract}

\maketitle


\section{Introduction}

There are several indirect pieces of observational evidence that outflows/jets are coupled to 
accretion disks in black hole accreting systems, from Galactic to extragalactic sizes. Recently, massive, highly ionized and mildly-relativistic Ultra-Fast Outflows (UFOs) have been detected in $>$40\% of radio-quiet Seyferts and quasars (e.g., Chartas et al.~2002,2003; Pounds et al.~2003; Markowitz et al.~2006; Braito et al.~2007; Cappi et al.~2009; Reeves et al.~2009; Tombesi et al.~2010a). They are essentially observed through blue-shifted Fe~XXV--XXVI absorption lines at E$>$7~keV.  
In particular, Tombesi et al.~(2011a) performed a photo-ionization modeling of these absorbers and derived the distribution of their main physical parameters. The outflow velocity is mildly-relativistic, in the range $\sim$0.03--0.3c, with a peak and mean value at $\sim$0.14c. The ionization is very high, in the range log$\xi$$\sim$3--6~erg~s$^{-1}$~cm, with a mean value of $\sim$4.2~erg~s$^{-1}$~cm. The column densities are large, in the interval $N_H$$\sim$$10^{22}$--$10^{24}$~cm$^{-2}$, with a mean value of $\sim$$10^{23}$~cm$^{-2}$. 
The characteristics of these UFOs are consistent with accretion disk winds/outflows and they can potentially play a significant role in the AGN cosmological feedback (e.g., King 2010).

In radio-loud AGNs, relativistic jets are routinely observed at radio, optical and X-rays. However, thanks to new long \emph{Suzaku} observations, evidence for accretion disk outflows in these sources is recently emerging. Here we report the discovery of UFOs with $v$$\sim$0.1c in 3/5 Broad-Line Radio Galaxies (BLRGs), namely 3C~111, 3C~120 and 3C~390.3 (Tombesi et al.~2010b; see \S~2), and the first evidence for an accretion disk-outflow connection in an AGN, 3C~111 (Tombesi et al.~2011b; see \S~3). 
The detection of UFOs in BLRGs represents an important step for models of jet formation and for our understanding of the jet-disk connection. From a theoretical perspective, disk outflows are a necessary, although not sufficient, ingredient for jet formation. For instance, in the magnetic tower jet simulations there is the possibility that some plasma is trapped and dragged upward by magnetic field lines and winds/outflows may provide the external pressure needed to effectively collimate the jet (e.g., Kato et al.~2004; McKinney~2006). Thus, models attempting to explain the link between the jet and the accretion process will have to take these components into account.

\begin{figure}
  \includegraphics[height=6.5cm, width=5.5cm]{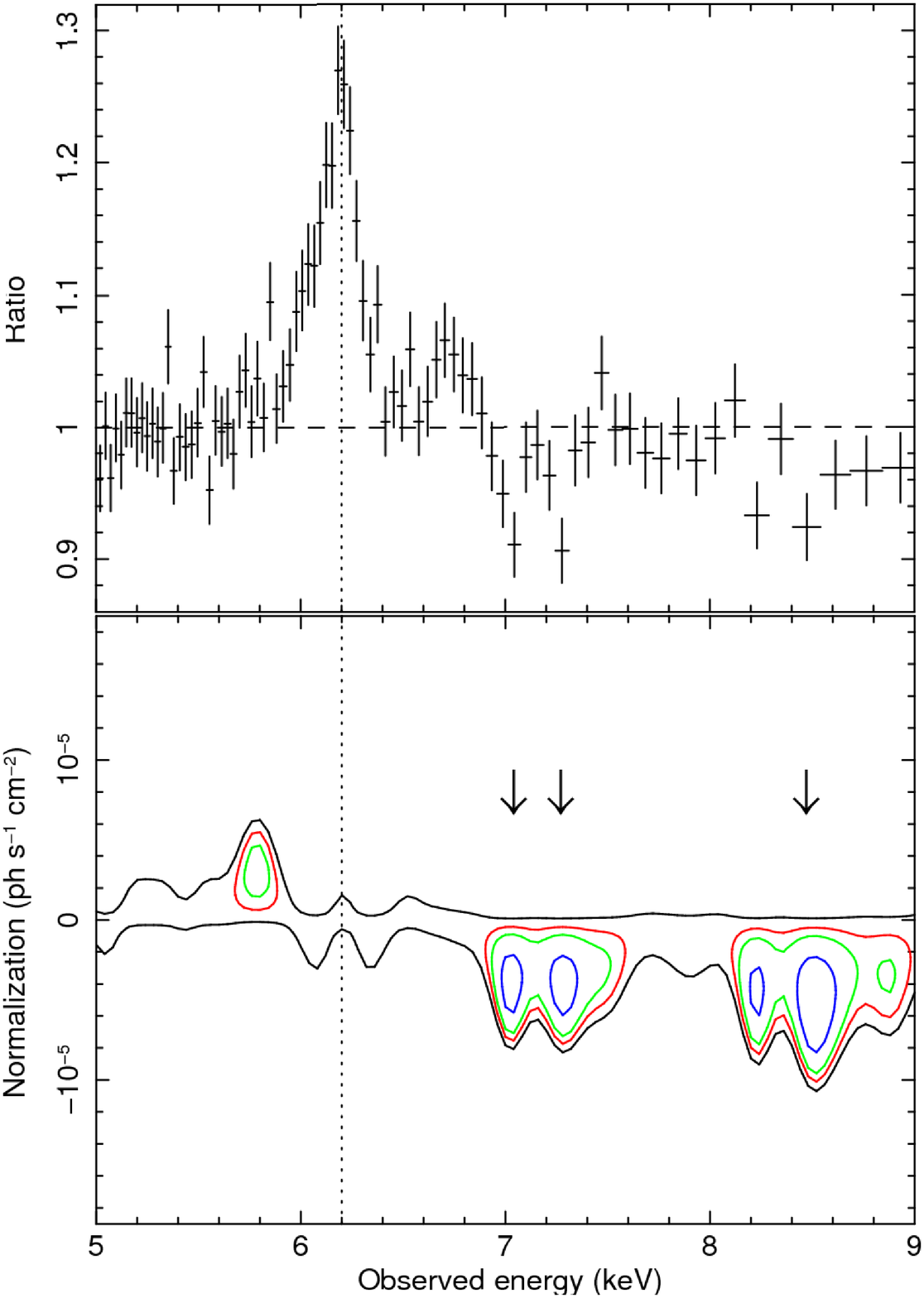}
\hspace{1cm}
  \includegraphics[height=6.5cm, width=5.5cm]{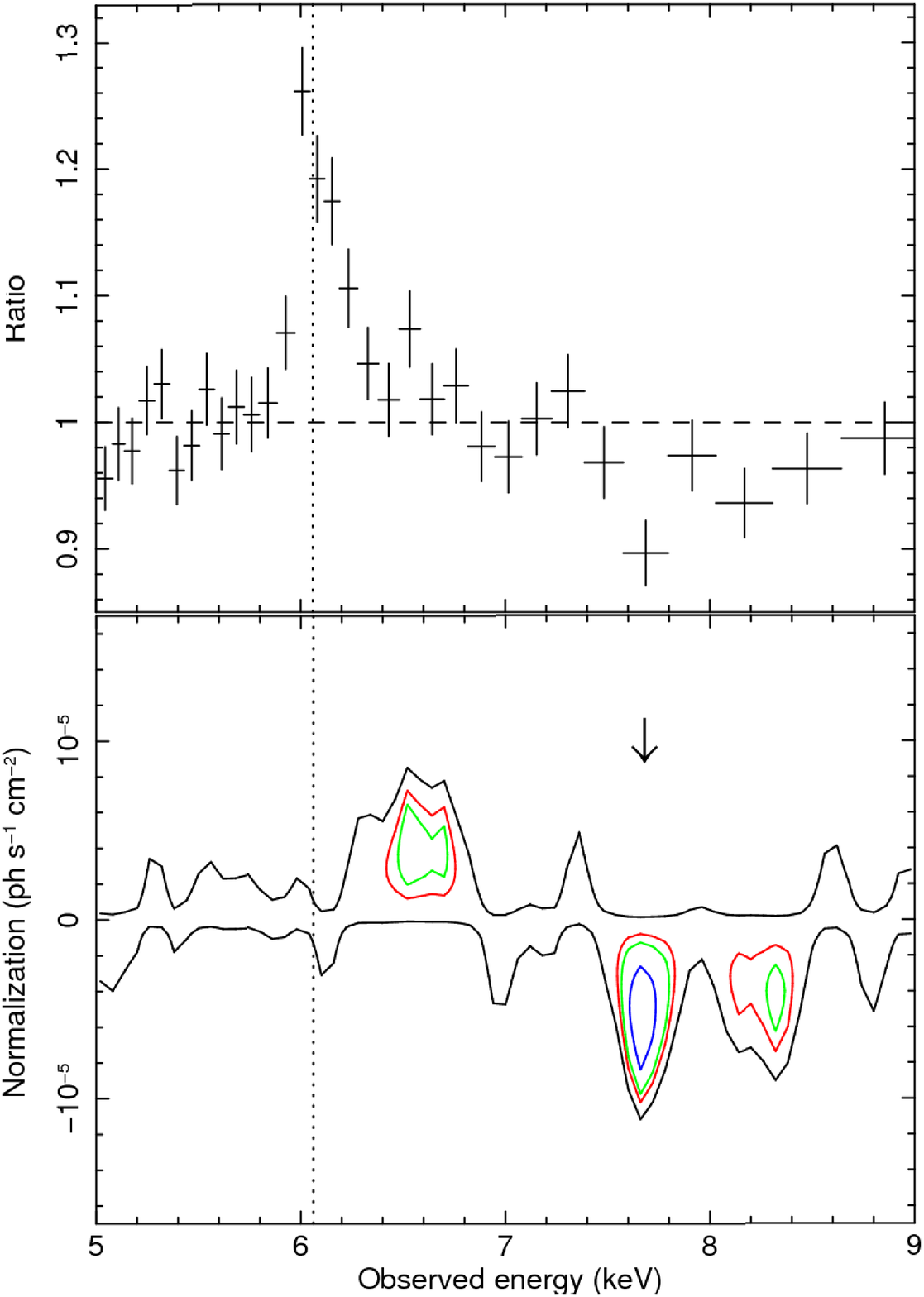}
  \caption{\emph{Suzaku} XIS spectra of 3C~120 (\emph{left}) and 3C~390.3 (\emph{right}) in the 5--9~keV band. \emph{Upper panel:} ratio against a power-law continuum. \emph{Lower panel:} energy-intensity contours with respect to a power-law continuum including also the Fe K emission lines. The contours refer to F-test confidence levels of 68\% (red), 90\% (green) and 99\% (blue), respectively. The arrows indicate the location of the blue-shifted absorption features.}
\end{figure}

\section{Ultra-Fast Outflows in BLRGs}

In Tombesi et al.~(2010b) we performed a systematic 4--10~keV spectral analysis of long \emph{Suzaku} observations of five BLRGs, i.e. 3C~111, 3C~120, 3C~390.3, 3C~382 and 3C~445. We detected, for the first time at X-rays in radio-loud AGNs, several absorption lines at energies greater than 7~keV in the first three sources. The lines are detected with high significance ($>$99\%) according to both the F-test and extensive Monte Carlo simulations (see Fig.~1). Their likely interpretation as blue-shifted Fe XXV--XXVI K-shell resonances implies an origin from highly ionized gas outflowing with mildly relativistic velocities, in the range $v$$\simeq$0.04--0.15c. A fit with specific photo-ionization models
gives ionization parameters in the range log$\xi$$\simeq$4--5.6~erg~s$^{-1}$~cm and column densities of $N_H$$\simeq$$10^{22}$--$10^{23}$~cm$^{-2}$. These characteristics are very similar to those of the UFOs observed in radio-quiet AGNs. Their estimated location within $\sim$0.01--0.1~pc from the central super-massive black hole suggests a likely origin related with accretion disk winds/outflows. The mass outflow rate of these UFOs is comparable to the accretion rate of $\sim$1~$M_{\odot}$~yr$^{-1}$. Their estimated kinetic or mechanical power is high, in the range $\sim$$10^{43}$--$10^{45}$~erg~s$^{-1}$, which is comparable to their typical jet power and corresponds to about their X-ray luminosity and $\sim$10\% of the bolometric luminosity.  Therefore, these UFOs can potentially play a significant role in the expected feedback from AGNs and can possibly be linked also to the jet activity.

\section{Accretion disk-wind connection in 3C~111}

\begin{figure*}
  \includegraphics[height=.3\textheight]{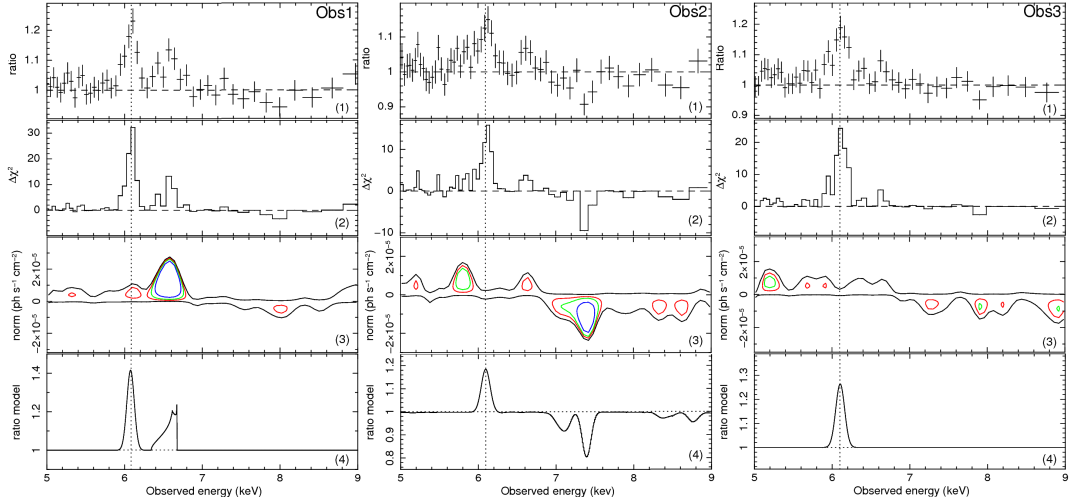}
  \caption{Comparison of the \emph{Suzaku} XIS spectra of the three new observations of 3C~111 in the 5--9~keV interval. \emph{Panel 1:} ratios with respect to a Galactic absorbed power-law continuum. \emph{Panel 2:} $\Delta \chi^2$ residuals. \emph{Panel 3:} contour plots with respect to the power-law continuum, including also the narrow Fe K$\alpha$ emission line. The contours refer to F-test confidence levels of 68\% (red), 90\% (green) and 99\% (blue), respectively. \emph{Panel 4:} final best-fitting models.}
\end{figure*}

The bright BLRG 3C~111 was proposed by us in the \emph{Suzaku} GO5 as the best candidate for a follow-up study of the predicted $\sim$7~days time-scale variability of its UFO (Tombesi et al.~2011b). It is important to note that 3C~111, as well as 3C~120, have been recently the subjects of an extensive monitoring campaign to study their accretion disk-jet connection (Chatterjee et al.~2009, 2011). 
The three \emph{Suzaku} observations have exposures of $\sim$60~ks and there is a $\sim$30\% flux increase between the first and the second. We performed a systematic 4--10~keV spectral analysis of the XIS data and derived that a Galactic absorbed power-law continuum plus a narrow Fe K$\alpha$ emission line at E$\simeq$6.4~keV provide a good initial representation of the data. However, an emission line at E$\simeq$6.88~keV is observed in the first observation and an absorption line at E$\simeq$7.75~keV in the second, see Fig.~2. The detection significance of these features is high, $>$99\% from both F-test and Monte Carlo simulations. They are also significantly variable among the observations.

However, the rest-frame energies of these lines are not directly consisntent with known atomic transitions. For the former, we find that a highly ionized relativistic line profile with the bulk of the emission coming from reflection off the accretion disk at $\sim$20--100$r_g$ from the black hole can provides a good modeling. For the latter, instead, we find that assuming the likely identification with blue-shifted Fe~XXV--XXVI absorption lines, a photo-ionization modeling suggest they are due to a highly ionoized, log$\xi$$\simeq$4.3~erg~s$^{-1}$~cm, UFO with outflow velocity $\sim$0.1c and column density $N_H$$\simeq$$8\times 10^{22}$~cm$^{-2}$. The location of this material is constrained at $<0.006$~pc from the black hole, which is consistent with accretion disk winds/outflows.

The mass outflow rate is comparable to the accretion rate of $\sim$1~$M_{\odot}$~yr$^{-1}$ and the mechanical power is $\sim$$5\times 10^{44}$~erg~s$^{-1}$, which is comparable to the typical jet power. This latter corresponds also to about half of the X-ray luminosity and $\sim$6\% of the bolometric luminosity. Therefore, this outflow constitutes a non negligible fraction of the overall energy budget of the AGN and can potentially provide a significant feedback effect besides the jet. 
In particular, this provides the first direct evidence for an accretion disk-wind connection in an AGN and is consistent with a picture in which an increased illumination of the inner disk causes an outflow to be lifted at $\sim$100~$r_g$. This is then possibly accelerated through radiation pressure to the observed terminal velocity of $\sim$0.1c, but additional magnetic thrust can not be excluded.

\end{document}